\documentclass[a4paper,11pt]{article}
\pdfoutput=1 

\usepackage{jheppub} 
\usepackage[english]{babel}
\usepackage[latin1]{inputenc}
\usepackage[caption=false]{subfig}
\usepackage{hyperref}
\newcommand{\be}{\begin{equation}}
\newcommand{\ee}{\end{equation}}
\newcommand{\ea}{\end{eqnarray}}
\newcommand{\ba}{\begin{eqnarray}}

\newcommand{\wt}{\widetilde}

\title{\boldmath Noncommutative approach to disclose a Higgs group}

\author{M. A. De Andrade}
\author{and C. Neves}
\affiliation{Departamento de Matem\'{a}tica, F\'{\i}sica e Computa\c{c}\~{a}o, Faculdade de Tecnologia, \\ Universidade do Estado do Rio de Janeiro,\\
Rodovia Presidente Dutra, Km 298, P\'{o}lo Industrial, CEP 27537-000, Resende-RJ, Brazil.}

\emailAdd{marco@fat.uerj.br}
\emailAdd{clifford@fat.uerj.br}

\abstract{A noncommutative(NC) version for a global $O(N)$ scalar field theory is proposed and an alternative  investigation about how noncommutative drives spontaneous symmetry breaking (SSB) is explored. Indeed, we show that the noncommutativity plays an important role in such mechanism, i.e., it is possible to show that there is a Higgs group with no more than two Higgs bosons. In this scenario, we establish two mutually exclusive options: one Higgs boson with mass at 125 GeV and other at 750 GeV -- 2 TeV  excess does not imply a 2 TeV mass resonance -- or two Higgs bosons with  mass-degenerate near 125~GeV, where 2 TeV and 750 GeV  excesses do not imply a 2 TeV and 750 GeV masses resonance.}

\begin{document}

\maketitle
\flushbottom

\section{Introduction}

\label{sec:intro}
Recently, we propose an alternative  method to induce noncommutativity into a commutative theory -- Noncommutative Mapping\cite{MACN}  --, where it was possible to setup different NC algebra with $\,2n(n-1)$ NC parameters into a $n$-dimensional system.  Therefore, it allowed us to explore different contributions related to the noncommutativity. This result driven us to generalize the $\ast$-product\cite{wigner,HG,MOYAL,mezincescu}. Further, it was also shown that different NC algebra among the phase-space coordinates origins different NC system and that the mass and charge are now NC parametrized. In another article\cite{MACN1}, it was shown that the NC parameter plays the role of the viscous damping coefficient in the damped harmonic oscillator(DHO) and, among other things,  the Noncommutative Mapping was applied in the global $O(N)$ scalar field theory, where the presence of damping feature was revealed and it was also discussed the relations among bosonic string attached to a 3D-brane, DHO, 2D-NC oscillator harmonic and  NC scalar field theory. More recently, we have been revealing\cite{MACN2} how spontaneous symmetry breaking (SSB) and Higgs-Kibble mechanism are driven by the noncommutativity and it was explored not only to explain, in an alternative way, the mass-degenerate Higgs bosons near 125 GeV, but also to see how the Higgs-Kibble mechanism changes in order to generate a NC dependent mass to the gauge fields.

CMS\cite{cms1,cms2,cms3} and ATLAS\cite{atlas1} collaboration have reported several excesses $\sim$2 TeV in the dijet invariant mass spectrum of $\sim$20.3 $fb^{-1}$ at $\sqrt{s}$ = 8TeV, for example: the ATLAS collaboration has reported that a 3.4$\sigma$, 2.6$\sigma$ and 2.9$\sigma$ deviation are observed $\sim$2 TeV in the invariant mass distribution of boosted WZ, WW
and ZZ, where the global significance of the discrepancy in the WZ channel is 2.5$\sigma$; the CMS experiment reported a moderate excess, $\sim$1.4$\sigma$ for the dijet resonances, where the W- and Z-tagged jets are indistinguishable; CMS
experiment reported a $\sim$ 2$\sigma$ excess at $\sim$1.8 TeV in the dijet resonance channel search. In this scenario, there are several papers\cite{Cacciapaglia,Carmona,Gao,Cao2,Cheung,Abe,Sanz,Thamm,Anchordoqui,Omura,Chen,Chiang,Dobrescu1,Brehmer,Dobrescu2,Hisano,Chao3} explaining this diboson excesses at 1.8 $\sim$2 TeV.

In 2015, with the LHC Run II -- ATLAS \cite{atlas2} and CMS\cite{cms4} --, an accumulated luminosity of $\sim$3 fb$^{-1}$ at $\sqrt{s}$=13~TeV showed a hint of a new particle at $\sim$750 GeV decaying into a photon pair. Despite of the 750 GeV excess may not involve a broad resonance with a mass near 750 GeV\cite{Cho,JBCS}, there are many ways to interpret the 750 GeV excess as being a 750 GeV mass resonance, for example: in the framework of a single new scalar particle\cite{Franceschini}, by singlets coupled to vector-like fermions \cite{butazzo,sdm,ellis,knapen,Kobakhidze,Falkowski,Chao1,Huang,Chakrabortty,Cao1,Dhuria,Blas,Murphy,Ding,Boucenna,Alves,Chao2}, composite states \cite{Harigaya,Low,Molinaro,No,Nakai,Bian,Chiara,Heckman,Matsuzaki,Hernandez}, reduction of extra dimensions \cite{Cox,Ahmed}, axions\cite{Pilaftsis,Higaki} or sgoldstinos\cite{Petersson,Demidov,Bellazzini}. Further, some authors start to explore a possible link of this new resonance to a dark matter particle \cite{Mambrini,Backovic,Barducci,Dev,Dey,Bi,Bauer}. Besides of all of this, we also find in the literature some articles where the authors assume that the 750 GeV diphoton excess is due to new Higgs boson(s) in Two-Higgs-Doublet Model (2HDM)\cite{georgi,Chanowitz,Branco,KKHL,Angelescu,Bertuzzo,Han,MB}.

Inspired by the 2HDM idea and by the NC contributions in mass-degenerate Higgs bosons\cite{MACN2}, we propose to disclose a Higgs boson group from the NC point of view. This work is organized as follows. In section \ref{sec:1}, we explore Noncommutative Mapping\cite{MACN} in field theory. In order to get this, a simple global $O(N)$ scalar theory is initially considered and, after that, we propose an \textit{ansatz} that allows us to get a particular NC version for $O(N)$ scalar field theory. In section \ref{sec:2}, a global $O(4)$ scalar field theory, with an internal symmetry group, is considered and, similarly to what was done in section \ref{sec:1}, a NC version field theory is obtained and the contribution of noncommutativity in the spontaneous symmetry breakdown mechanism\cite{nambu,goldstone,lasinio1,lasinio2,higgs1,higgs2,kibble,gsw,IZ,TL,kaku,ryder} might be properly explored: we show that there is a Higgs group with only two Higgs bosons, where they can be interpreted as being the one with mass equal to 125 GeV and the other with 750 GeV -- there is no room to accomodate the 2TeV excess -- or two Higgs bosons with  mass-degenerate near 125~GeV, where 2 TeV and 750 GeV  excesses do not imply a 2 TeV and 750 GeV masses resonance. At the end, some conclusions are presented.
\section{NC scalar field theory}
\label{sec:1}

In order to investigate the contribution of noncommutativity in the context of field theory, a simplest scalar field in four space-time dimensions is considered, namely, a global $O(N)$ scalar field theory, whose its dynamics is governed by
\be
\label{lp42}
{\cal{L}}=\frac 12 (\partial_\mu\phi_i)(\partial^\mu\phi_i)-\frac{\mu^2}{2}\phi_i\phi_i-\frac {\lambda}{4} (\phi_i\phi_i)^2 ,
\ee
where $\lambda$ is a positive number, $\mu^2$ can be either positive or negative and the field $\phi_i$ transforms as an $N$-vector. The corresponding Hamiltonian is
\be
\label{hamiltoniana21}
{\cal{H}}=\frac{\pi_i\pi_i}{2}+\frac{\nabla\phi_i\nabla\phi_i}{2} + U,
\ee
with the following potential
\be
\label{Cpotential}
U=\frac{\mu^2}{2}\phi_i\phi_i+\frac {\lambda}{4} (\phi_i\phi_i)^2 .
\ee
It is well know that if $\mu^2>0$, then the vacuum is at $\phi_i\phi_i=0$ and the symmetry is manifest, and $\mu^2$ is the mass of the scalar modes. On the other hand, if $\mu^2<0$, there is a new vacuum solution given by $\phi_i\phi_i={\frac{-{\mu}^2}{\lambda}}$, which has an infinite number of possible vacua. 

In the commutative framework, the symplectic variables are $\xi^\beta=(\phi_i,\pi_i)$ and the symplectic matrix is
\be
f=\left(
\begin{array}{cc}
 0           ~&~   \delta_{ij} \cr
 -\delta_{ij}   ~&~  0 \cr
\end{array}
\right)\delta^{(3)}(\boldsymbol{x}-\boldsymbol{y}) .
\ee
The noncommutativity is introduced into the model changing the brackets among the phase-space variables, given by
\be
\label{ST045aaa}
\left\{\wt {\phi}_i, \wt {\phi}_j\right\} = 0,\,\,
\left\{\wt {\phi}_i, \wt{\pi}_j\right\} = \delta_{ij}\delta^{(3)}(\boldsymbol{x}-\boldsymbol{y}),\,\,
\left\{\wt{\pi}_i, \wt{\pi}_j\right\} = \Theta_{ij}\delta^{(3)}(\boldsymbol{x}-\boldsymbol{y}),
\ee
where the time-dependent antisymmetric quantity, $\Theta_{ij}$, embraces the noncommutativity. These brackets are comprised by the symplectic matrix in NC basis, namely:
\be
\label{ST050aaa}
{\wt{f}}=\left(
\begin{array}{cc}
 0 ~&~    \delta_{ij}  \cr
 -\delta_{ij}     ~&~   \Theta_{ij}\cr
\end{array}
\right)\delta^{(3)}(\boldsymbol{x}-\boldsymbol{y}).
\ee
The NC transformation matrix\cite{MACN}, $~R=\sqrt{\wt{f}\,f^{-1}}~$, is written as
\be
\label{ST062aaa}
R=\left(
\begin{array}{cc}
 \delta_{ij}             ~&~   0  \cr
 \frac12\,\Theta_{ij}    ~&~  \delta_{ij} \cr
\end{array}
\right)\delta^{(3)}(\boldsymbol{x}-\boldsymbol{y}) .
\ee
Since the commutative symplectic variables $~\xi^\beta=(\phi_i,\pi_i)$ change to the NC ones $\wt\xi^\alpha=(\wt{\phi}_i, \wt{\pi}_i)$ through $d\wt\xi^\alpha=R^{\alpha}_{~\beta}\,d\xi^\beta~,$ it follows that
\ba
\label{q-trans1a}
\wt{\phi}_i&=&{\phi}_i~,\,\,
\wt{\pi}_i=\pi_i +\frac12\,\Theta_{ij}\,{\phi}_j~.
\ea
In agreement with the NC Mapping\cite{MACN} the NC first-order Lagrangian can be read as
\be
\label{ST066a}
\wt{\cal{L}}(\phi_i,\dot{\phi}_i)=\pi_i\,\dot{\phi}_i-{\wt{\cal{H}}}(\phi_i,\pi_i),
\ee
where $\wt{\cal{H}}(\phi_i,\pi_i)={\cal{H}}(\wt{\phi}_i,\wt{\pi}_i)$ and the latter one is the NC version of the Hamiltonian, Eq.(\ref{hamiltoniana21}), given by
\be
\label{ST067}
{\cal{H}}(\wt{\phi}_i,\wt{\pi}_i)=\frac{\wt{\pi}_i\wt{\pi}_i}{2} +\frac{\nabla\wt\phi_i\nabla\wt\phi_i}{2}+\frac{\mu^2}{2} \wt{\phi}_i\wt{\phi}_i+ \frac{\lambda}{4}(\wt{\phi}_i\wt{\phi}_i)^2.
\ee
The Hamiltonian density above, with the help of Eq.(\ref{q-trans1a}), renders to
\be
\label{hamiltonianalp41a}
\wt{\cal{H}}(\phi_i,\pi_i)=\frac{\pi_i\pi_i}{2} + \frac 12\pi_i\Theta_{ij}\phi_j+\frac {\nabla\phi_i\nabla\phi_i}{2} + \wt{U},
\ee
where 
\be
\label{potential2a}
\wt{U}=\frac{\mu^2}{2}\phi_i\phi_i+\frac{\lambda}{4}(\phi_i\phi_i)^2+\frac18\Theta_{ij}\Theta_{ik}\phi_j\phi_k.
\ee
Observe that the original model is restored when $\Theta_{ij}$ is a null quantity.  Occasionally, energy density might be written as being the sum of kinetic and potential energy\cite{coleman},
\be
E=T+V,
\ee
where, in the Eq.(\ref{hamiltonianalp41a}), $T$ is the two first term and $V$, as usual, is the term involving no time derivatives, namely,
\be
V=\frac{\nabla\phi_i\nabla\phi_i}{2}+\wt{U}.
\ee
As a consequence, if the energy is to be bounded below, $\wt{U}$ must be also bounded below. 

The Hamilton's equation of motion $\left(\dot\phi_i=\frac{\partial \wt{\cal{H}}(\phi_i,\pi_i)}{\partial\pi_i}\right)$ is calculated and the canonical momenta is obtained as being
\be
\label{HEquationa}
\pi_i= \dot\phi_i-\frac 12\Theta_{ij}\phi_j.
\ee
Inserting Eq.(\ref{hamiltonianalp41a}), with Eq.(\ref{potential2a}), and Eq.(\ref{HEquationa}) into the NC first-order Lagrangian in Eq.(\ref{ST066a}), we get the NC second-order Lagrangian
\be
\label{Covlagrangiana}
\wt{\cal{L}}=\frac 12 (\partial_\mu\phi_i)(\partial^\mu\phi_i) -\frac 12 (n^\mu\partial_\mu\phi_i)\Theta_{ij}\phi_j-\frac{\mu^2}{2}\phi_i\phi_i-\frac{\lambda}{4}(\phi_i\phi_i)^2 ,
\ee
with the time-like vector $n^\mu= (1,\boldsymbol{0})$, which is a normal vector of a noncovariant set of equitemporal surfaces ($t$ = constant) where the Hamiltonian analysis is implemented. However, this noncovariance is apparent, because if we consider a larger set of space-like surfaces to develop the Hamiltonian formalism, this obstruction can be removed\footnote{This observation is well clarified by one of us in the appendix A of Ref.\cite{WN}}. From this point of view, $\Theta_{ij}$ appear as a set of Lagrange multipliers that imposes the velocity dependent constraint $(\partial_\mu\phi_i)\phi_j$. As pointed out by some authors\cite{BST,BHZ,WuZee}, a Lagrangian, first-order in velocity $(\dot\phi_i)$, can always be considered as arising from a $U(1)$ background potential in configuration space. At this point, we would like to point out that the middle term of the right hand side of this NC Lagrangian plays the role of damped term\cite{MACN1}.

In order to investigate how the noncommutativity drives the spontaneous symmetry breaking and Higgs-Kibble mechanism, we assume the dimension of the internal group as being $N=2^n,\texttt{ with } n\in \mathbb{N}^+$,  and consider the following \textit{ansatz}, 
\be
\label{teta}
\Theta_{ij}=\theta\Sigma_{ij},
\ee
where $\theta$ could be a constant or a time-dependent parameter, and $\Sigma_{ij}$ are the elements of a constant antisymmetric matrix $\Sigma$.
For $N=2$, we have $\Sigma=\varepsilon$, where $\varepsilon$ is the $2\times2$ antisymmetric matrix with $\varepsilon_{12}=1$ and, consequently, the \textit{ansatz} renders to
\be
\label{teta2}
\Theta_{ij}=\theta\varepsilon_{ij}.
\ee
For $N>2$, the $\Sigma$ matrix is given by
\be
\Sigma=\left(
\begin{array}{ccccc}
 \Gamma ~&~ 0~&~ \dots ~&~0 \cr
0 ~&~  \Gamma~&~ \dots ~&~0 \cr
\vdots ~&~\vdots ~&~ \ddots ~&~\vdots\cr
0 ~&~ 0 ~&~ \dots ~&~\Gamma\cr
\end{array}
\right),
\ee
where
\be
\Gamma=\frac1{\sqrt{2}}\left(
\begin{array}{cc}
 \varepsilon~&~   \varepsilon\cr
 \varepsilon~&~  -\varepsilon \cr
\end{array}
\right).
\ee
Due to this, we get
\be
\label{epsilon}
\Sigma_{ik}\Sigma_{kj}=-\delta_{ij},
\ee
with $i,j=1,2,\dots,N$. At this moment, it is important to point out that there are alternative choices for the NC $\Theta_{ij}$-parameter, Eq.(\ref{teta}), and that each choice generates a different result, which gives room to explore new features.

Implementing the result, given in Eq.(\ref{epsilon}), into the potential, given in Eq.(\ref{potential2a}), we get
\be
\label{potential3bb}
\wt{U}=\frac{\wt{\mu}^2}{2}\phi_i\phi_i+\frac {\lambda}{4}(\phi_i\phi_i)^2,
\ee
with $\wt{\mu}^2=\mu^2+\frac14\theta^2$. Note that the original mass can now be tuned by the NC $\theta$-parameter. 

Further, we can also consider, from the beginning, that $\mu^2=0$. In this scenario, the NC potential, Eq.(\ref{potential3bb}), renders to
\be
\label{potential3}
\wt{U}=\frac{\theta^2}{4}\phi_i\phi_i+\frac{\lambda}{4}(\phi_i\phi_i)^2,
\ee
while the NC Lagrangian, Eq.(\ref{Covlagrangiana}), reduces to
\be
\label{Covlagrangian01}
\wt{\cal{L}}=\frac 12 (\partial_\mu\phi_i)(\partial^\mu\phi_i)-\frac 12 n^\mu\partial_\mu\phi_i\Theta_{ij}\phi_j-\frac{\lambda}{4}(\phi_i\phi_i)^2.
\ee
At this point, we would like to point out that $\theta^2$ is a positive definite parameter from the beginning, \textit{i.e.}, $\theta^2>0$.

\section{Spontaneous symmetry breaking}
\label{sec:2}
Let us now examine a simple example given by the global O(4) scalar theory with an internal symmetry group, where $\chi_i^\alpha$ transform as a 4-vector and $\alpha=1,\,2$:
\be
\label{ssb0010}
{\cal{L}}=\frac 12 \partial_\mu\chi_i^\alpha\partial^\mu\chi_i^\alpha-\frac 12\mu^{2}\chi_i^\alpha\chi_i^\alpha-\frac{\lambda}{4}(\chi_i^\alpha\chi_i^\alpha)^2,
\ee
where $\lambda$ is a positive number and $\mu^2$ can be either positive or negative. The corresponding Hamiltonian is
\be
\label{hamiltoniana21a}
{\cal{H}}=\frac{\pi_i^\alpha\pi^\alpha_i}{2}+\frac{\nabla\chi_i^\alpha\nabla\chi_i^\alpha}{2} + U,
\ee
with the following potential
\be
\label{Cpotentialabc}
U=\frac{\mu^2}{2}\chi^\alpha_i\chi^\alpha_i+\frac {\lambda}{4} (\chi_i^\alpha\chi_i^\alpha)^2 .
\ee 

In the commutative framework, the symplectic variables are $\xi=(\chi_i^\alpha,\pi^\alpha_i)$ and the symplectic matrix is
\be
f=\left(
\begin{array}{cc}
 0           ~&~   \delta_{ij} \cr
 -\delta_{ij}   ~&~  0 \cr
\end{array}
\right)\delta^{(3)}(\boldsymbol{x}-\boldsymbol{y})\delta^{\alpha\beta} .
\ee
The noncommutativity is introduced into the model changing the brackets among the phase-space variables, given by
\be
\label{ST045aab}
\left\{\wt {\chi}^\alpha_i, \wt {\chi}^\beta_j\right\} = 0,\,\,
\left\{\wt {\chi}^\alpha_i, \wt{\pi}^\beta_j\right\} = \delta_{ij}\delta^{\alpha\beta}\delta^{(3)}(\boldsymbol{x}-\boldsymbol{y}),\,\,
\left\{\wt{\pi}^\alpha_i, \wt{\pi}^\beta_j\right\} = \Theta_{ij}\delta^{\alpha\beta}\delta^{(3)}(\boldsymbol{x}-\boldsymbol{y}),
\ee
where antisymmetric matrix, $\Theta_{ij}$, embraces the noncommutativity. These brackets are comprised by the symplectic matrix in NC basis, namely:

\be
\label{ST050aab}
{\wt{f}}=\left(
\begin{array}{cc}
 0 ~&~    \delta_{ij}  \cr
 -\delta_{ij}     ~&~   \Theta_{ij}\cr
\end{array}
\right)\delta^{\alpha\beta}\delta^{(3)}(\boldsymbol{x}-\boldsymbol{y}).
\ee
The NC transformation matrix\cite{MACN}, $~R=\sqrt{\wt{f}\,f^{-1}}~$, is written as
\be
\label{ST062aab}
R=\left(
\begin{array}{cc}
 \delta_{ij}             ~&~   0  \cr
 \frac12\,\Theta_{ij}    ~&~  \delta_{ij} \cr
\end{array}
\right)\delta^{\alpha\beta}\delta^{(3)}(\boldsymbol{x}-\boldsymbol{y}) .
\ee
For $\alpha=\beta$, the matrix $\Theta$ is given by
\ba
\label{ST06234}
\Theta&=&\frac{\theta}{\sqrt{2}}
\left(
\begin{array}{cc}
 \varepsilon  ~&~   \varepsilon  \cr
 \varepsilon   ~&~  -\varepsilon \cr
\end{array}
\right),\nonumber\\
\Theta&=&{\theta}\,\Gamma
\ea
where
\be
\label{ST06234a}
\Gamma=\frac{1}{\sqrt{2}}\left(
\begin{array}{cc}
 \varepsilon  ~&~   \varepsilon  \cr
 \varepsilon   ~&~  -\varepsilon \cr
\end{array}
\right),
\ee
and
\be
\label{ST06234aa}
\varepsilon=\left(
\begin{array}{cc}
0&1\cr
-1&0\cr
\end{array}
\right).
\ee

Since the commutative symplectic variables $~\xi=(\chi_i^\alpha,\pi^\alpha_i)$ change to the NC ones $\wt\xi=(\wt{\chi}^\alpha_i, \wt{\pi}^\alpha_i)$ through $\wt\xi=R\,\xi~,$ it follows that
\ba
\label{q-trans1}
\wt{\chi}^\alpha_i&=&{\chi}^\alpha_i~,\,\,
\wt{\pi}^\alpha_i=\pi^\alpha_i +\frac{\theta}{2}\Gamma_{ij}\,{\chi}^\alpha_j~.
\ea

In agreement with the NC Mapping\cite{MACN} the NC first-order Lagrangian can be read as
\be
\label{ST066}
\wt{\cal{L}}(\chi^\alpha_i,\dot{\chi}^\alpha_i)=\pi^\alpha_i\,\dot{\chi}^\alpha_i-{\wt{\cal{H}}}(\chi^\alpha_i,\pi^\alpha_i)~.
\ee
As $\wt{\cal{H}}(\chi^\alpha_i,\pi^\alpha_i)={\cal{H}}(\wt{\chi}^\alpha_i,\wt{\pi}^\alpha_i)$ and ${\cal{H}}(\wt{\chi}^\alpha_i,\wt{\pi}^\alpha_i)$ is the NC version of the Hamiltonian, Eq.(\ref{hamiltoniana21a}), becomes
\be
\label{ST067a}
{\cal{H}}(\wt{\chi}^\alpha_i,\wt{\pi}^\alpha_i)=\frac{\wt{\pi}^\alpha_i\wt{\pi}^\alpha_i}{2} +\frac{\nabla\wt\chi^{\alpha}_i\nabla\wt\chi^{\alpha}_i}{2}+\frac{\mu^{2}}{2} \wt{\chi}^\alpha_i\wt{\chi}^\alpha_i+ \frac{\lambda}{4}(\wt{\chi}^\alpha_i\wt{\chi}^\alpha_i)^2
\ee
the Hamiltonian density $\wt{\cal{H}}(\chi^\alpha_i,\pi^\alpha_i)$, with the help of Eq.(\ref{q-trans1}), renders to

\be
\label{hamiltonianalp41}
\wt{\cal{H}}(\chi^\alpha_i,\pi^\alpha_i)=\frac{\pi^\alpha_i\pi^\alpha_i}{2} + \frac{\theta}{2}\pi^\alpha_i\Gamma_{ij}\chi^\alpha_j+\frac {\nabla\chi_i^{\alpha}\nabla\chi_i^{\alpha}}{2} + \wt{U},
\ee
where 
\be
\label{potential2}
\wt{U}=\frac{\wt{\mu}^{2}}{2}\chi^\alpha_i\chi^\alpha_i+\frac{\lambda}{4}(\chi^\alpha_i\chi^\alpha_i)^2,
\ee
with
\be
\label{tmass}
\wt{\mu}^{2}={\mu}^{2}+\frac{\theta^2}{4}.
\ee

The Hamilton's equation of motion $\left(\dot\chi^\alpha_i=\frac{\partial \wt{\cal{H}}(\chi^\alpha_i,\pi^\alpha_i)}{\partial\pi^\alpha_i}\right)$ is calculated and the canonical momenta is obtained as being
\be
\label{HEquation}
\pi^\alpha_i= \dot\chi^\alpha_i-\frac{\theta}{2}\Gamma_{ij}\chi^\alpha_j.
\ee
Inserting Eq.(\ref{hamiltonianalp41}), with Eq.(\ref{tmass}), and Eq.(\ref{HEquation}) into the field version of the NC Lagrangian given in Eq.(\ref{ST066}), the NC second-order Lagrangian is obtained, namely
\be
\label{Covlagrangian}
\wt{\cal{L}}=\frac 12 (\partial_\mu\chi^\alpha_i)(\partial^\mu\chi^\alpha_i) -\frac {\theta}{2} \dot{\chi}^\alpha_i\Gamma_{ij}\chi^\alpha_j-\frac{\mu^{2}}{2}\chi^\alpha_i\chi^\alpha_i-\frac{\lambda}{4}(\chi^\alpha_i\chi^\alpha_i)^2 ,
\ee

Since each two scalar fields can be combined into each single complex scalar field, we write
\ba
\label{ssb0020}
\phi^\alpha_1&=&\frac{1}{\sqrt{2}}(\chi_1^\alpha+\imath\chi_2^\alpha),\nonumber\\
\phi^{\alpha\ast}_1&=&\frac{1}{\sqrt{2}}(\chi_1^\alpha-\imath\chi_2^\alpha),\nonumber\\
\phi^\alpha_2&=&\frac{1}{\sqrt{2}}(\chi_3^\alpha+\imath\chi_4^\alpha),\\
\phi^{\alpha\ast}_2&=&\frac{1}{\sqrt{2}}(\chi_3^\alpha-\imath\chi_3^\alpha),\nonumber
\ea
where 
\be
\label{ssb00030}
\phi^\alpha=\left(
\begin{array}{c}
\phi^\alpha_1\cr
\phi^\alpha_2\cr
\end{array}
\right)
\ee
is a doublet representation of $SU(2)\otimes SU(2)$ with an internal symmetry group. The Lagrangian, given in Eq.(\ref{Covlagrangian}), renders to
\be
\label{Covlagrangian10}
\wt{\cal{L}}=(\partial_\mu\phi^{\alpha\dagger})(\partial^\mu\phi^{\alpha})-{\imath\theta}\dot{\phi}^{\alpha\dagger}\Gamma\phi^{\alpha}-\mu^{2}\phi^{\alpha\dagger}\phi^{\alpha}-\lambda(\phi^{\alpha\dagger}\phi^{\alpha})^2,
\ee
and applying the usual Legendre transformation, the Hamiltonian is computed and it is given by
\be
\wt{\cal H}=\pi^{\alpha\dagger}\pi^\alpha+(\nabla\phi^{\alpha\dagger})\nabla\phi^\alpha-\imath\theta ({\pi^\alpha}^T\phi^\alpha-{\pi^\alpha}^\dagger{\phi^\alpha}^\ast)+\wt{U},
\ee
where the potential $\wt{U}$ is 
\be
\label{potentialUNC}
\wt{U}=\wt{\mu}^{2}\phi^{\alpha\dagger}\phi^\alpha+\lambda(\phi^{\alpha\dagger}\phi^{\alpha})^2.
\ee
The potential is minimal at
\ba
\label{vacuum}
\left\langle \phi^{\alpha\dagger}\phi^{\alpha} \right\rangle_0 &=&-\frac{\wt{\mu}^{2}}{2\lambda},\nonumber\\
\left\langle \phi^{\alpha\dagger}\phi^{\alpha} \right\rangle_0 &=&\frac{v^2}{2},\\
\left\langle \phi^{\alpha\ast}_1\phi^{\alpha}_1+\phi^{\alpha\ast}_2\phi^{\alpha}_2  \right\rangle_0 &=&\frac{{v}^2}{2},\nonumber
\ea
where
\be
\label{v2}
{v}^2=-\frac{\wt{\mu}^2}{\lambda}.
\ee
A possible solution is
\ba
\label{ssb23}
\left\langle  \phi^{\alpha\ast}_1\phi^\alpha_1\right\rangle_0 &=&0\nonumber\\
\left\langle \phi^{\alpha\ast}_2\phi^\alpha_2 \right\rangle_0 &=&\frac{{v}^2}{2}
\ea
Then, the field $\phi^\alpha_1=(\phi_1^1\,\,\,\phi_1^2)$ has a charge $Q^\alpha_1=1$, while $\phi^\alpha_2$ has a null charge, \textit{i.e.}, $Q^\alpha_2=0$. From the second equation given in Eq.(\ref{ssb23}), we get
\ba
\label{ssb123}
\left\langle \phi^{1\ast}_2\phi^1_2 +\phi^{2\ast}_2\phi^2_2\right\rangle_0 &=&\frac{{v}^2}{2},\nonumber\\
(v_2^1)^2+(v_2^2)^2=v^2&=&\frac{-\mu^2-\frac{\theta^2}{4}}{\lambda},\\
{\lambda}(v_2^1)^2+{\lambda}(v_2^2)^2&=& {-\mu^2-\frac{\theta^2}{4}}.\nonumber
\ea
An \textit{educated guess} solution is
\ba
\label{higgsMass}
{\lambda}(v_2^1)^2&=&-\frac{\mu^2}{2}-\theta^2,\nonumber\\
{\lambda}(v_2^2)^2&=&-\frac{\mu^2}{2}+\frac{3\theta^2}{4}.
\ea
At this point, we would like to point out that the internal symmetry spontaneously breakdown, in an analogous way to what it happens to $SU(2)\otimes SU(2)$ symmetry group.

The doublet $\phi^\alpha$ obeys the $SU(2)\otimes SU(2)$ transformation property
\be
\phi^\alpha\rightarrow e^{\imath g \tau^\alpha_i\eta^\alpha_i/2}\left(\phi^\alpha+\frac{v^\alpha}{\sqrt{2}}\right)
\ee
with
\be
v^\alpha=
\left(
\begin{array}{c}
0 \cr
v_2^\alpha
\end{array}
\right),
\ee
as a constant $SU(2)\otimes SU(2)$ doublet. 

We can reparametrize $\phi^\alpha$ in the following way:
\be
\label{phiparametrized}
\phi^\alpha\rightarrow \frac{1}{\sqrt{2}}e^{\imath g \tau^\alpha_i\zeta^\alpha_i/v}\left(
\begin{array}{c}
0\cr
v_2^\alpha+H^\alpha
\end{array}
\right)
\ee
with $\zeta^\alpha_i(x)$ and $H(x)$ as real fields and where $\tau^\alpha_i$ are the SU(2) group generators. With this parametrization it is apparent that the fields $\zeta^\alpha_i(x)$ can be transformed always by a $SU(2)\otimes SU(2)$ gauge transformation, as in the case of the Abelian Higgs model. This choice, which is called the unitary gauge, is perfectly adequate for calculations in the semi-classical limit. However, it must be abandoned beyond this limit. Here we will set $\zeta^\alpha_i(x)$ = 0 and, consequently, the reparametrization above reduces to
\be
\label{tranp1}
\phi^\alpha\rightarrow \frac{1}{\sqrt{2}}\left(
\begin{array}{c}
0\cr
v_2^\alpha+H^\alpha
\end{array}
\right)
\ee
Inserting the equation above into Eq.(\ref{potentialUNC}), we get
\ba
\label{potentialUNC2}
\wt{U}&=&\wt\mu^2\left[\frac{1}{{2}}\left(
\begin{array}{cc}
0 & v_2^\alpha+H^\alpha
\end{array}
\right)
\left(
\begin{array}{c}
0 \cr
v_2^\alpha+H^\alpha
\end{array}
\right)
\right]+\lambda\left[\frac{1}{{2}}\left(
\begin{array}{cc}
0 & v_2^\alpha+H^\alpha
\end{array}
\right)
\left(
\begin{array}{c}
0 \cr
v_2^\alpha+H^\alpha
\end{array}
\right)
\right]^2,\nonumber\\
&=&\frac{\wt\mu^2}{{2}}\left[(v_2^{\alpha})^2+(H^{\alpha})^2+2v_2^{\alpha}H^{\alpha}\right]+\frac{\lambda}{{4}}\left[(v_2^{\alpha})^2+(H^{\alpha})^2+2v_2^{\alpha}H^{\alpha}\right]^2,\nonumber\\
&=&\frac{1}{{2}}\left[\wt\mu^2+3\lambda (v_2^\alpha)^2\right](H^{\alpha})^2+\dots\nonumber\\
&=&\frac{1}{{2}}\left[\wt\mu^2+3\lambda (v_2^1)^2\right](H^{1})^2+\frac 12 \left[\wt\mu^2+3\lambda (v_2^2)^2\right](H^{2})^2+\dots.\nonumber\\
\ea
Inserting Eq.(\ref{higgsMass}) into the equation above, we get
\be
\wt{U}=\frac{1}{{2}}\left[-\frac{\mu^2}{2}-\frac{11}{2}\theta^2\right](H^{1})^2+\frac 12 \left[-\frac{\mu^2}{2}+\frac 52\theta^2\right](H^{2})^2+\dots.\nonumber\\
\ee
which allows us to infer that the Higgs scalar doublet $H^\alpha$ has a squared masses:
\ba
\label{higgsmass1}
m^2_{H^1}&=&-\frac{\mu^2}{2}-\frac{11}{4}\theta^2=M^2-\frac{11}{4}\theta^2,\nonumber\\
m^2_{H^2}&=&-\frac{\mu^2}{2}+\frac 52\theta^2=M^2+\frac 52\theta^2,
\ea
where $M^2=-\frac{\mu^2}{2}$, $\mu^2<0$ and  $\theta^2>0$. Here, $H^1$ and  $H^2$ can be interpreted, respectively, as being the Higgs boson with mass equal to 125 GeV and 750 GeV.

Now, consider another \textit{educated guess} solution for Eq.(\ref{ssb123}), given by
\ba
\label{vacuum1}
{\lambda}(v_2^1)^2&=&-\frac{\mu^2}{2}-\frac 16 \theta^2,\nonumber\\
{\lambda}(v_2^2)^2&=&-\frac{\mu^2}{2}-\frac {1}{12} \theta^2.
\ea
the Higgs scalar doublet $H^\alpha$ has a squared masses:
\ba
\label{higgsmass1234}
m^2_{H^1}&=&-\frac{\mu^2}{2}-\frac{1}{4}\theta^2=M^2-\frac{1}{4}\theta^2,\nonumber\\
m^2_{H^2}&=&-\frac{\mu^2}{2}=M^2.
\ea
This result can be interpreted in a two distinct and mutually exclusive way: first, $H^2$ can be interpreted as being the Higgs boson with mass equal to 125 GeV and $\theta$ can be settle in order to get $H^1$ as being the Higgs boson near 125 GeV, i.e., the mass-degenerate Higgs bosons near 125 GeV\cite{MACN2,MH2,MH3,MH4,MH5,MH6,MH7} can be explained and the 750 GeV excess does not imply a 750 GeV mass resonance; second, $H^2$ can be interpreted as being the Higgs boson with mass equal to 750 GeV and $\theta$ can be settle in order to get $H^1$ as being the Higgs boson at 125 GeV.

Another \textit{educated guess} solution for Eq.(\ref{ssb123}) is
\ba
\label{vacuum1234}
{\lambda}(v_2^1)^2&=&-\frac{\mu^2}{2}-\frac 18 \theta^2,\nonumber\\
{\lambda}(v_2^2)^2&=&-\frac{\mu^2}{2}-\frac {1}{8} \theta^2.
\ea
This hypothesis allows us to infer that the Higgs scalar doublet $H^\alpha$ has the same following squared masses, given by
\be
\label{higgsmass2}
m^2_{H^1}=m^2_{H^2}=-\frac{\mu^2}{2}-\frac 18 \theta^2=M^2-\frac 18 \theta^2.\\
\ee
Here, $H^1$ and  $H^2$ can be interpreted as being two Higgs boson with 125 GeV and, due to some kind of an interaction among them, the mass-degenerate Higgs bosons near 125 GeV\cite{MACN2,MH2,MH3,MH4,MH5,MH6,MH7} might appear.

Further, we can also propose a general solution for Eq.(\ref{ssb123}), 
\ba
\label{higgsMass5a}
{\lambda}(v_2^1)^2&=&-x_1{\mu^2}+y_1{\theta^2},\nonumber\\
{\lambda}(v_2^2)^2&=&-x_2{\mu^2}+y_2{\theta^2},
\ea
with
\ba
\label{restrina1}
x_1+x_2&=&1,\nonumber\\
y_1+y_2&=&-\frac 14.
\ea
Inserting Eq.(\ref{higgsMass5a}) into Eq.(\ref{potentialUNC2}), we get
\be
\wt{U}=\frac{1}{{2}}\left[(1-3x_1){\mu^2}+\left(\frac 14+ 3y_1\right){\theta^2}\right](H^{1})^2+\frac 12 \left[(1-3x_2){\mu^2}+\left(\frac 14+3y_2\right){\theta^2}\right](H^{2})^2+\dots
\ee
The coefficients of $H^\alpha$ must be non-null and positive, then 
\ba
\label{restrain7a}
&(&1-3x_\alpha)\mu^2+\left(\frac{1}{4}+3y_\alpha\right)\theta^2>0,\nonumber\\
x_\alpha&>&\frac 13 -\left(\frac 14 + 3y_\alpha\right)\frac{\theta^2}{3|\mu^2|} \texttt{ with }\alpha=1,\,2.
\ea 
The sum of the two equations above is given by
\be
\label{restrain8a}
x_1+x_2>\frac 23 - \frac{\theta^2}{6|\mu^2|}-\frac{\theta^2}{\mu^2|}(y_1+y_2)
\ee 
Inserting Eq.(\ref{restrina1}) into the equation above, we get
\ba
1&>&\frac 23 - \frac{\theta^2}{6|\mu^2|} + \frac{\theta^2}{4|\mu^2|},\nonumber\\
\frac 13&>&  \frac{\theta^2}{12|\mu^2|},\nonumber\\
|\mu^2|&>&\frac{\theta^2}{4}.
\ea
Note that the Higgs group has two fields and there is a restrain between the NC $\theta$-parameter and the mass mode$(\mu)$.

Another possible discussion arises when the index group $\alpha$ is enlarged, i.e., $\alpha=1,\,2,\,3$. In this context, Eq.(\ref{ssb123}) changes to
\ba
\label{ssb1234}
\left\langle \phi^{1\ast}_2\phi^1_2 +\phi^{2\ast}_2\phi^2_2+\phi^{3\ast}_2\phi^3_2\right\rangle_0 &=&\frac{{v}^2}{2},\nonumber\\
(v_2^1)^2+(v_2^2)^2+(v_2^3)^2=v^2&=&\frac{-\mu^2-\frac{\theta^2}{4}}{\lambda},\\
{\lambda}(v_2^1)^2+{\lambda}(v_2^2)^2+{\lambda}(v_2^3)^2&=& {-\mu^2-\frac{\theta^2}{4}},\nonumber
\ea
and the potential, Eq.(\ref{potentialUNC2}), reduces to
\be
\label{potentialUNC23}
\wt{U}=\frac{1}{{2}}\left[\wt\mu^2+3\lambda (v_2^1)^2\right](H^{1})^2+\frac 12 \left[\wt\mu^2+3\lambda (v_2^2)^2\right](H^{2})^2+\frac 12 \left[\wt\mu^2+3\lambda (v_3^2)^2\right](H^{3})^2+\dots.
\ee

An \textit{educated guess} solution for Eq.(\ref{ssb1234}) is
\ba
\label{higgsMass5}
{\lambda}(v_2^1)^2&=&-x_1{\mu^2}-\theta^2,\nonumber\\
{\lambda}(v_2^2)^2&=&-x_2{\mu^2}+\frac{\theta^2}{4},\\
{\lambda}(v_2^3)^2&=&-x_3{\mu^2}+\frac{\theta^2}{2},\nonumber
\ea
with
\be
\label{restrina}
x_1+x_2+x_3=1.
\ee
Inserting these solution on Eq.(\ref{potentialUNC23}), we get
\ba
\label{potentialUNC24}
\wt{U}&=&\frac{1}{{2}}\left[\wt\mu^2-3x_1{\mu^2}-3\theta^2\right](H^{1})^2+\frac 12 \left[\wt\mu^2-3x_2{\mu^2}+\frac 34\theta^2\right](H^{2})^2+\frac 12 \left[\wt\mu^2-3x_3{\mu^2}+\frac 32\theta^2\right](H^{3})^2+\dots,\nonumber\\
&=&\frac{1}{{2}}\left[(+1-3x_1){\mu^2}-\frac{11}{4}\theta^2\right](H^{1})^2+\frac 12 \left[(+1-3x_2){\mu^2}+\theta^2\right](H^{2})^2\nonumber\\
&+&\frac 12 \left[(+1-3x_3){\mu^2}+\frac{7}{4}\theta^2\right](H^{3})^2+\dots.
\ea
The coefficients of $\mu^2$ must be non-null and negative, then the parameters $x_1,\,x_2,\,x_3$ is bounded below, $x_1,\,x_2,\,x_3>\frac 13\Rightarrow x_1+x_2+x_3>1$. This constrain together to the relation given in Eq.(\ref{restrina}) lead us to conclude that the solution set for $x_1,\,x_2,\,x_3$ is the null set. Therefore, the symmetry does not spontaneously break and, consequently, it is not possible to have a doublet representation, $SU(2)\otimes SU(2)$ with an internal symmetry group, $\phi_i^\alpha$ with $\alpha=1,\,2,\,3$, which embraces three  Higgs bosons, at least in the NC approach. The later procedure can be applied in a doublet representation, $SU(2)\otimes SU(2)$ with an internal symmetry group, $\phi_i^\alpha$ with $\alpha>3$ and, in an analogously way to what was done for $\phi_i^\alpha$ with $\alpha=1,\,2,\,3$, we can conclude that the spontaneous symmetry breaking mechanism is obliterated, consequently, it does not exist a Higgs group with more than two Higgs bosons.

On the other hand, a general \textit{educated guess} solution for Eq.(\ref{ssb1234}) is
\ba
\label{higgsMass6}
{\lambda}(v_2^1)^2&=&-x_1{\mu^2}+y_1\theta^2,\nonumber\\
{\lambda}(v_2^2)^2&=&-x_2{\mu^2}+y_2{\theta^2},\\
{\lambda}(v_2^3)^2&=&-x_3{\mu^2}+y_3{\theta^2},\nonumber
\ea
with
\ba
\label{restrina7}
x_1+x_2+x_3=1,\nonumber\\
y_1+y_2+y_3=-\frac 14.
\ea
Inserting these solution on Eq.(\ref{potentialUNC23}), we get
\ba
\label{potentialUNC24}
\wt{U}&=&\frac{1}{{2}}\left[\wt\mu^2-3x_1{\mu^2}+3y_1\theta^2\right](H^{1})^2+\frac 12 \left[\wt\mu^2-3x_2{\mu^2}+ 3y_2\theta^2\right](H^{2})^2+\frac 12 \left[\wt\mu^2-3x_3{\mu^2}+3y_3\theta^2\right](H^{3})^2+\dots,\nonumber\\
&=&\frac{1}{{2}}\left[(+1-3x_1){\mu^2}+\left(\frac{1}{4}+3y_1\right)\theta^2\right](H^{1})^2+\frac 12 \left[(+1-3x_2){\mu^2}+\left(\frac{1}{4}+3y_2\right)\theta^2\right](H^{2})^2\nonumber\\
&+&\frac 12 \left[(+1-3x_3){\mu^2}+\left(\frac{1}{4}+3y_3\right)\theta^2\right](H^{3})^2+\dots.
\ea
The coefficients of $H^\alpha$ must be non-null and positive, then 
\ba
\label{restrain8}
&(&1-3x_\alpha)\mu^2+\left(\frac{1}{4}+3y_\alpha\right)\theta^2>0,\nonumber\\
x_\alpha&>&\frac{\theta^2/|\mu^2|+4}{12}+\frac{\theta^2}{|\mu^2|}y_\alpha \texttt{ with }\alpha=1,\,2,\,3.
\ea
The sum of the three equations above is given by
\be
\label{restrain9}
x_1+x_2+x_3>\frac{\theta^2/|\mu^2|+4}{4}+\frac{\theta^2}{|\mu^2|}\left(y_1+ y_2+ y_3\right).
\ee 
Inserting Eq.(\ref{restrina7}) into the equation above, we get the following statement: $1>1$. Therefore, the existence of a Higgs group with a third field is not possible. If the previous procedure is applied, in an analogous way what was done for $\alpha=1,\,2,\,3$, in a model with $\alpha>3$ the same result is obtained and, consequently, it was demonstrated, from a NC point of view, that there is no room to accommodate  in the Higgs group more than two Higgs fields.

At this point, we would like to stress that the contribution of noncommutativity into the Higgs-Kibble mechanism, which is VEV dependent, was investigated in a previous work\cite{MACN2}.

\section{Conclusion}
\label{sec:3}
We would like to point out that the NC $\theta$-parameter affects the spontaneous symmetry breaking mechanism, \textit{vide} section \ref{sec:2}, in an astonishing way due to the doublet representation of $SU(2)\otimes SU(2)$ with an internal symmetry group, $\phi_i^\alpha,\,\,\alpha=1,\,2$. Here, it was revealed that, when the spontaneous symmetry is breakdown, the NC $\theta$-parameter changes the energy vacuum such that $\phi_i^\alpha$ can be reparametrized, \textit{vide} Eq.(\ref{tranp1}), which drives us to establish the following conclusion: for a doublet representation of $SU(2)\otimes SU(2)$ with an internal symmetry group, where $\alpha>2$, the spontaneous symmetry breakdown mechanism is obliterated and, consequently, there is a Higgs group with only two Higgs bosons. In this scenario, we argue that the 2 TeV  excess does not imply a 2 TeV mass resonance and, also, we can interpret these two Higgs bosons, with a NC dependent mass, in the following way: (1) from Eq.(\ref{higgsmass1}) we get two Higgs bosons, one at 125 GeV and other at 750 Gev; (2) from Eq.(\ref{higgsmass1234}) and Eq.(\ref{higgsmass2}) the Higgs group presents mass-degenerate Higgs bosons near 125~GeV and, consequently, 750 GeV  excess does not imply a 750 GeV mass resonance.

\section*{Acknowledgments}
 
C. Neves and M. A. De Andrade thank to Brazilian Research Agencies(CNPq and FAPERJ) for partial financial support. Further, we would like to thank Bruno Fernando Inchausp Teixeira for a careful reading of the manuscript.

\end{document}